# Modular and Cost-effective Scanning Photocurrent Microscopy System for Sub-micron characterization of 2D optoelectronic devices

Nuria Jiménez-Arévalo [1,†,*], Dan Zheng [2,3,4; †], Yong Xie [1], Peng Cheng [2,3,4], Yi Liu [2,3,4], Rafael Luque Merino [1], Haiwen Yu [2,3,4], Tao Wang [2,3,4], Riccardo Frisenda [5], Andres Castellanos-Gomez [1,*], Qinghua Zhao [2,3,4,*],

1. 2D Foundry Research Group, Instituto de Ciencia de Materiales de Madrid (ICMM-CSIC), Madrid, 28049, Spain
2. State Key Laboratory of Solidification Processing, Northwestern Polytechnical University, Xi'an, 710072, P. R. China
3. Key Laboratory of Radiation Detection Materials and Devices, Ministry of Industry and Information Technology, Xi'an, 710072, P. R. China
4. Research &Development Institute of Northwestern Polytechnical University in Shenzhen, Shenzhen, 518063, P. R. China
5. Dipartimento di Fisica, Università di Roma "La Sapienza", 00185, Rome, Italy

† Both authors contributed equally
* Corresponding authors:
nuria.jimenez.arevalo@csic.es
andres.castellanos@csic.es
qinghua_zhao@nwpu.edu.cn

## Abstract

Scanning photocurrent microscopy (SPCM) is a powerful technique for probing local optoelectronic phenomena in 2D semiconducting devices. However, commercial setups remain costly, complex and often lack flexibility and adaptability. In this work, we present a home-built SPCM platform built around the retrofitting of a conventional metallographic microscope by coupling it with different light sources (single-mode fiber-coupled lasers and multimode fiber-coupled high-power LEDs), a motorized XY stage, a digital camera and an electronic readout module. This system enables simultaneous acquisition of photocurrent and reflection intensity maps, requiring minimal modifications of the microscope. We reached sub-micron spatial resolution and high imaging fidelity by correlating photocurrent maps with reflection maps, optical micrographs and AFM topography data on different devices fabricated with different materials (InSe, $MoS_2$, $WSe_2$, Gr), on different substrates ($Si/SiO_2$, compact disk). This work provides a reliable, accessible and reproducible high-performance SPCM platform that can be easily implemented in most laboratories for microscale optoelectronic characterization of 2D devices.

## Introduction

Since the isolation of graphene in 2004, two-dimensional (2D) materials have attracted significant attention owing to their wide range of electrical, optical, and mechanical properties [1–5]. Many semiconducting 2D materials exhibit strong light-matter interaction, broadband absorption and high carrier mobility [6–10], making them promising candidates for next-generation optoelectronic devices, including photodetectors, phototransistors, photodiodes and other light imaging devices [6,7,11–13] . The performance of such devices, characterized by responsivity, response time or quantum efficiency [14–17], is directly linked to the microscopic mechanisms governing photocurrent generation and charge transport [18,19].

Conventional photocurrent characterizations are typically performed under global illumination, providing averaged optoelectronic device characteristics[20]. However, this is a limited approach, as it does not capture spatially varying phenomena such as Schottky barriers, local built-in fields, defects, strain gradients or layer-dependent absorption. To address this limitation, spatially localized techniques such as light-beam induced current (LBIC) mapping [21–24] or scanning photocurrent microscopy (SPCM) have been developed. In particular, SPCM has proven to be effective for studying 2D materials [12,24–27], enabling direct visualization of interfacial band alignment[28] , junction behaviour[12,24,29], depletion regions [24,30], photothermal effects [31–33] and defect-mediated transport at the microscale[34,35].

Although the usefulness of SPCM is well established [36], its widespread adoption is still limited. Commercial systems are scarce, expensive, and technically complex, and often lack the flexibility required for customization, thereby limiting accessibility and innovation. Previous reports demonstrated that home-built SPCM setups can drastically reduce costs and offer greater adaptability and accessibility [37,38]. However, such systems are usually limited to spatial resolutions on the order of a few micrometers, can be difficult to reproduce, or do not provide simultaneous photocurrent and reflection imaging. As a result, SPCM remains inaccessible to many laboratories hindering future advances in 2D optoelectronic devices.

To address this need, we present a fully modular, cost-effective and reproducible SPCM platform that can be easily built by the modification of a standard metallographic microscope, using commercially available optical, mechanical and electronic components. Its modular design is entirely based on standard threaded optical elements, allowing straightforward reconfiguration of the illumination source, spot size and optical fiber geometry without modifying the microscope body. This system integrates fiber-coupled light sources (single-mode lasers or high-power LEDs) and a motorized XY stage, allowing for simultaneously acquisition of photocurrent and reflection images. We demonstrate sub-micron spatial resolution for some configurations and validate the setup through detailed reflection and photocurrent mapping on $MoS_2$, InSe and $WSe_2$ devices. We highlight its capability to resolve Schottky barriers, bias-dependent carrier separation and wavelength-dependent absorption. The presented SPCM system aims to spread the use of this high-resolution optoelectronic characterization technique across different laboratories working on 2D materials and nanoscale devices.

## Results and discussion

Figure 1a shows a schematic of the home-built scanning photocurrent microscopy (SPCM) system developed for spatially resolved optoelectronic characterization of 2D-based devices, highlighting the different components. The setup is based on a modified Motic BA310 MET-H metallurgical microscope (see Figure 1b), adapted to allow flexible control of the optical path and maximize the illumination intensity delivered to the sample. To this end, the binocular viewing port was removed while keeping the internal tube lens intact. A custom-machined adapter was used to connect the original dovetail mount to standard SM1-threaded optomechanical components (Thorlabs). A 90:10 cube beamsplitter was installed immediately after the tube lens. The 90% transmission path was directed toward the fiber-coupled illumination input, while the 10% reflection path was routed to a 0.92 megapixels digital camera (USB output), enabling real-time imaging and intensity monitoring.

The fiber input path includes a focusing stage (SM1NR1, Thorlabs) and a XY translation stage (ST1XY-S, Thorlabs) for lateral alignment. These allow precise placement of the fiber core at the image plane of the objective lens system and centering it in the field of view of the camera. The correct positioning is achieved by visual inspection. After focusing the sample under white light illumination, the white light is turned OFF and an ND filter is inserted in front of the camera to prevent saturation from the intense spot. The fiber-coupled light source is then turned ON and the fiber core is brought into focus until a sharp image is projected onto the sample surface. For multimode fibers and incoherent sources (e.g., LEDs), this results in a clearly defined circular spot; for single-mode lasers, some interference fringes may be observed (see Figure S3).

The sample is mounted on a motorized XY stage (75 × 75 mm travel range), controlled via USB, which provides submicron step resolution (0.125 μm) and maximum scan speeds up to 10 mm/s. Manual XY-θ coarse alignment is available via an additional base stage (see Figure 1b). The electrical measurements are performed using a source-measure unit (SMU, Keithley 2450) connected to the sample via two miniature electrical probes. As shown in Figure 1c, these probes are mounted on XYZ micromanipulators (7T25-10XYZ, Standa) for precise tip positioning. The sample is placed on a temperature-controlled heating platform, enabling optional temperature regulation during measurements. Each probe incorporates magnetic bases to ensure firm contact and mechanical stability. The flexible cantilevered electrical tips are adapted from a commercial SD card connector, providing compliant vertical contact with the sample while preserving an ultra-low profile (Figure 1d). This design facilitates operation under high-magnification objectives with restricted working distance. Additional details on the design and fabrication of the probe assembly are provided in the Supporting Information.

An important feature of our system is its ability to acquire photocurrent and reflectance maps simultaneously. Unlike many conventional SPCM approaches that typically estimate the scanned region by acquiring two images (at the (0, 0) and (end, end) coordinates), our setup captures an image at every scan point. The laser or LED spot, attenuated by a neutral density (ND) filter, is imaged by the camera, and its RGB intensity is recorded at each pixel. This approach provides a real-time reflection map, directly co-registered with the photocurrent signal. Beyond enabling precise spatial mapping correlation between device features and photocurrent spots, this strategy also offers intrinsic diagnostics to detect motor malfunctions (e.g., stage getting stuck), thermal drifts, or dynamic changes in the sample (e.g., electrostatic discharge or oxidation) during acquisition, issues that would go unnoticed in conventional methods. Furthermore, this configuration eliminates the need for a separate photodetector and additional beam splitter, preserving light intensity and reducing hardware complexity.

Additional implementation details (including all component part numbers, supplier references, and custom-designed elements) are provided in Tables S1–S2 and Figures S1–S2 of the Supporting Information.

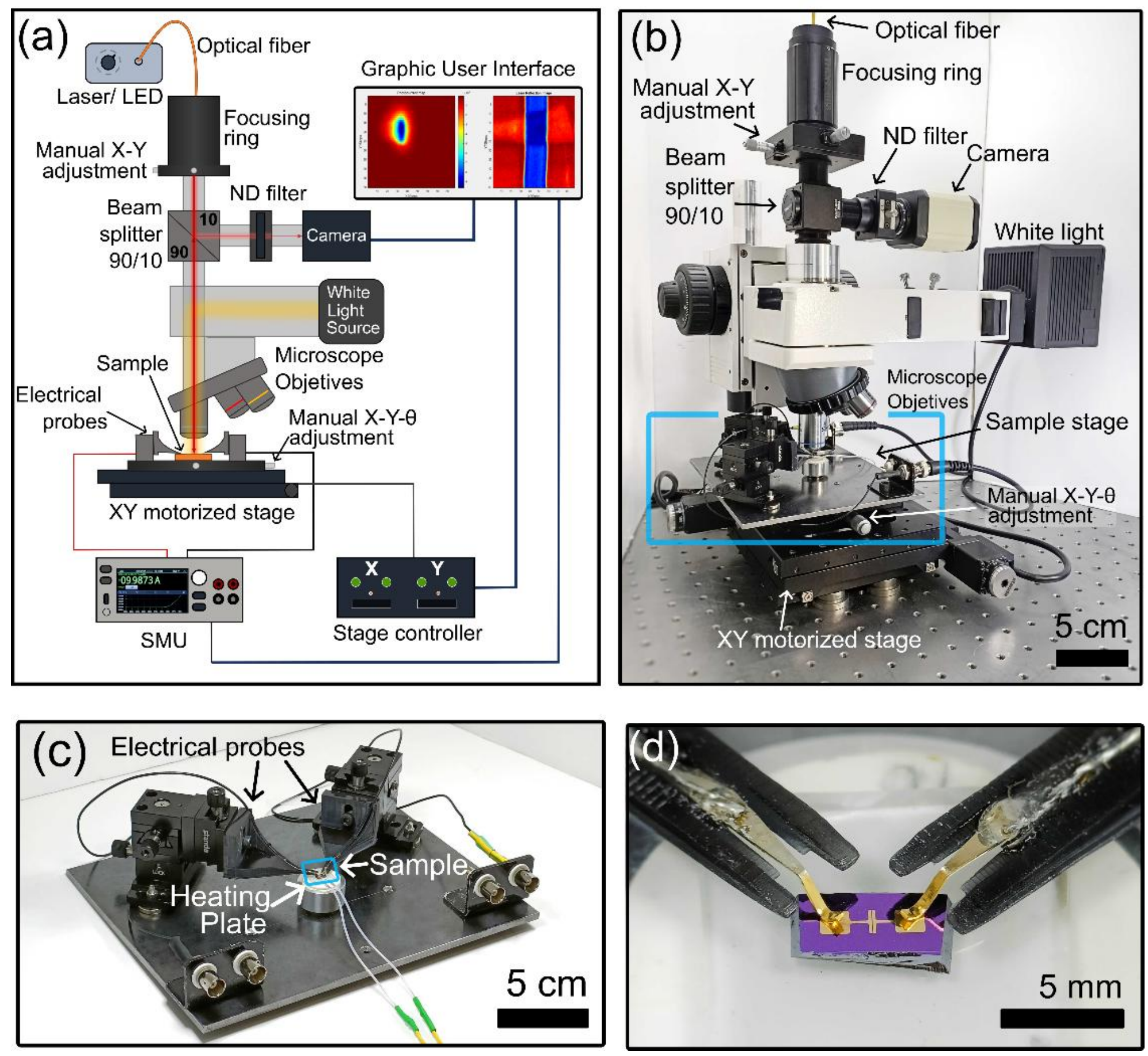

*Figure 1.* (a) Schematic diagram of the scanning photocurrent microscopy (SPCM) setup, showing the main optical and electrical components and connections, and representative data outputs displayed in the graphical user interface. (b) Photograph of the modified metallurgical microscope platform, highlighting the main optical elements and positioning system. (c) Photograph of the custom-built sample stage with integrated heating plate and miniature electrical probes (blue box in (b)). (d) Close-up photograph of a mounted device (blue box in (c)).

In the following, we describe the operation of the SPCM system. Focusing the optical fibre core onto the sample creates a micron-scale (or even submicron, depending on the used microscope magnification and optical fiber) illumination spot. This localized excitation can generate electron-hole pairs in semiconducting devices or induce local photothermal effects in materials such as graphene or metals, both of which can lead to measurable photocurrent signals[32,33]. Upon external biasing, or the presence of local built-in potentials (e.g. Schottky barriers, PN junctions, etc.), these photogenerated carriers can result in a measurable photocurrent. By raster-scanning the focused light beam across the sample surface, one can construct a spatially resolved photocurrent and reflectance distribution maps, providing insights into the structure and performance of 2D optoelectronic devices [19,24,33] .

As already mentioned, a key advantage of this type of SPCM setup is its ability to establish a precise correspondence between micrometric morphological features and the electrical signal produced upon illumination. In devices based on 2D-material, typical surface features, such as flake/electrode contacts, bubbles, wrinkles, or terraces with different numbers of layers are well known to influence the performance of devices[39–44]. To assess these characteristics, we used our home-built scanning photocurrent system to examine surface features of InSe flakes deposited on Si/$SiO_2$ substrates with Pt electrodes.

Figure 2 (a) shows an optical microscopy image of the edge of a InSe flake deposited onto a Si/$SiO_2$ substrate. The right side corresponds to bare Si/$SiO_2$ and the left side is covered with a InSe flake. Figure 2b displays the spatial map of reflected light intensity at 650 nm, collected as the laser beam raster-scans over the region shown in Figure 2a with a 0.5 μm step size for both the X and Y directions. The reflected intensity varies across the surface, with a stronger signal on the InSe flake as compared to the Si/$SiO_2$ substrate (see also Figure S3 and the discussion there) due to different interference conditions. Additionally, atomic force microscopy (AFM) topography was acquired in the same region (Figure 2c). Two cross-sectional line profiles from Figures 2b and 2c were extracted, and plotted them as a function of distance in Figure 2d.

To further evaluate the capability of the system to resolve fine topographical details, Figures 2e-g highlight its performance in scanning device features. Figure 2e shows an optical microscopy image of a Pt-InSe-Pt device, revealing visible bubbles in the InSe nanoflake, likely caused by a combination of adsorbate trapping during the flake transfer and air-induced passivation upon atmospheric exposure[45,46]. Figure 2f presents the reflection intensity map of the micro-region indicated by the red square in Figure 2e, clearly demonstrating the correlation between the very small optical features (~2 μm diameter bubbles) and the scanned reflection data. Similarly, Figure 2g compares the optical image of a wrinkle in the InSe nanoflake on a $Si/SiO_2$ substrate with its corresponding reflection intensity map. In both cases, the reflection intensity distributions closely match the optical images, demonstrating the high resolution of the system and accuracy in distinguishing key structural features. Further demonstration of the good spatial resolution of this SPCM, including data for a $MoS_2$ flake deposited on a compact disk (CD) substrate, can be found in section 4 of the Supporting Information. The CD substrate was chosen because it provides surface features on the ~1.5 μm scale, enabling resolution benchmarking.

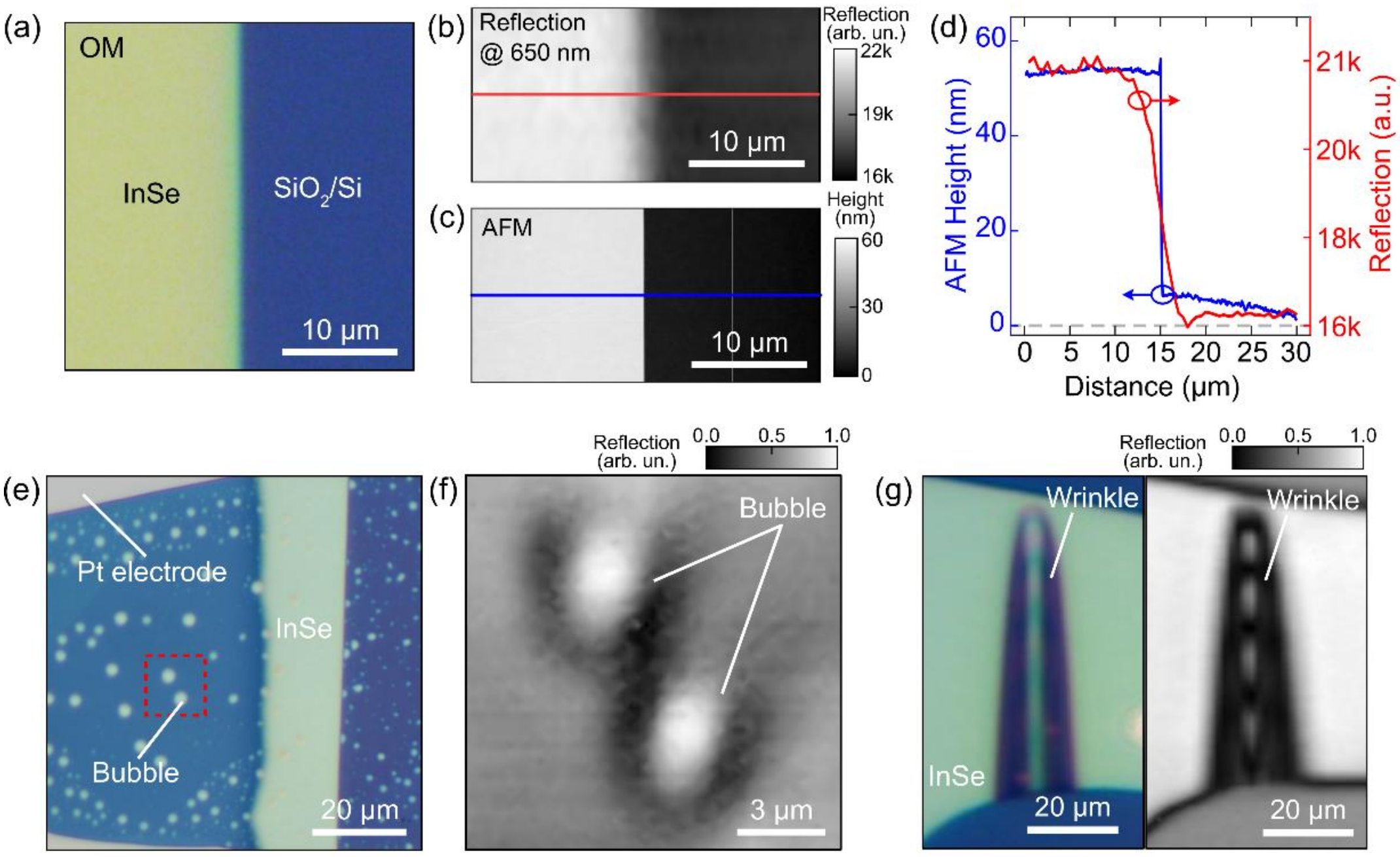

*Figure 2.* Spatially resolved laser (650 nm) reflection mapping using the SPCM system to probe surface morphology. (a) Optical microscopy (OM), (b) laser reflection map, and (c) atomic force microscopy (AFM) of an InSe nanoflake on a $Si/SiO_2$ substrate. (d) Cross-sectional profiles of AFM height and laser reflection intensity along the scan lines indicated in panels (b) and (c) showing a step from InSe to the substrate. (e) Optical image of an InSe flake featuring trapped bubbles. (f) Reflection map of two of the bubbles, corresponding to the region marked in panel (e). (g) Optical microscopy image (left) and corresponding reflection map (right) of an InSe wrinkle.

Next, we illustrate the capability of spatially correlate photocurrent with topographic features in a four-layer $MoS_2$-based device (see Figure S6). Figure 3(a) shows a grayscale map of the reflected light intensity across the scanned device area. The Pt electrodes appear brighter due to their higher reflectivity compared to $MoS_2$ and $Si/SiO_2$. The dashed outlines mark materials boundaries, providing a reference to align and interpret the spatially resolved photocurrent data. Figures 3 (b-d) display the local photocurrents measured as the laser spot scans across the device under different applied drain (D) -source (S) voltages ($V_{DS}$). The scanning step size was 0.3 μm, with one current measurement per step, and 0.5 s intervals between steps. In total we recorded current and reflection maps of 4350 pixels in less than 40 minutes.

At each metal (Pt)-semiconductor ($MoS_2$) interface, the built-in electric field associated to the Schottky barrier drives photo-generated electrons and holes in opposite directions. As the two contacts form a back-to-back Schottky junction, the direction of the built-in field reverses from one electrode to the other. As a result, illumination near one contact produces a photocurrent of one polarity, while illumination near the opposite contact generates a photocurrent of the opposite polarity.

At $V_{DS}$ = 0.0 V (Figure 3c), a bipolar photocurrent pattern is observed, with a negative response near the Pt(D)-$MoS_2$ interface and a positive one near the $MoS_2$-Pt(S) interface. This behaviour originates from the opposite orientations of the built-in fields at the two Schottky contacts, indicating an asymmetric band alignment [24].

When an external bias is applied, the asymmetry between the two Schottky contacts increases, and only one of the photocurrents hot-spots remains visible. Under negative bias ($V_{DS}$ = -0.5 V; Figure 3b), the drain-side Schottky barrier increases, resulting in an increase of the negative photocurrent in that contact. In contrast, the built-in field in the source is strongly suppressed, so no measurable photocurrent is present on that contact.

On the contrary, under positive bias ($V_{DS}$ = +0.5 V; Figure 3d), the source-side Schottky barrier increases while the drain-side barrier is suppressed. As a consequence, only the positive photocurrent lobe in the source contact is present in the photocurrent map.

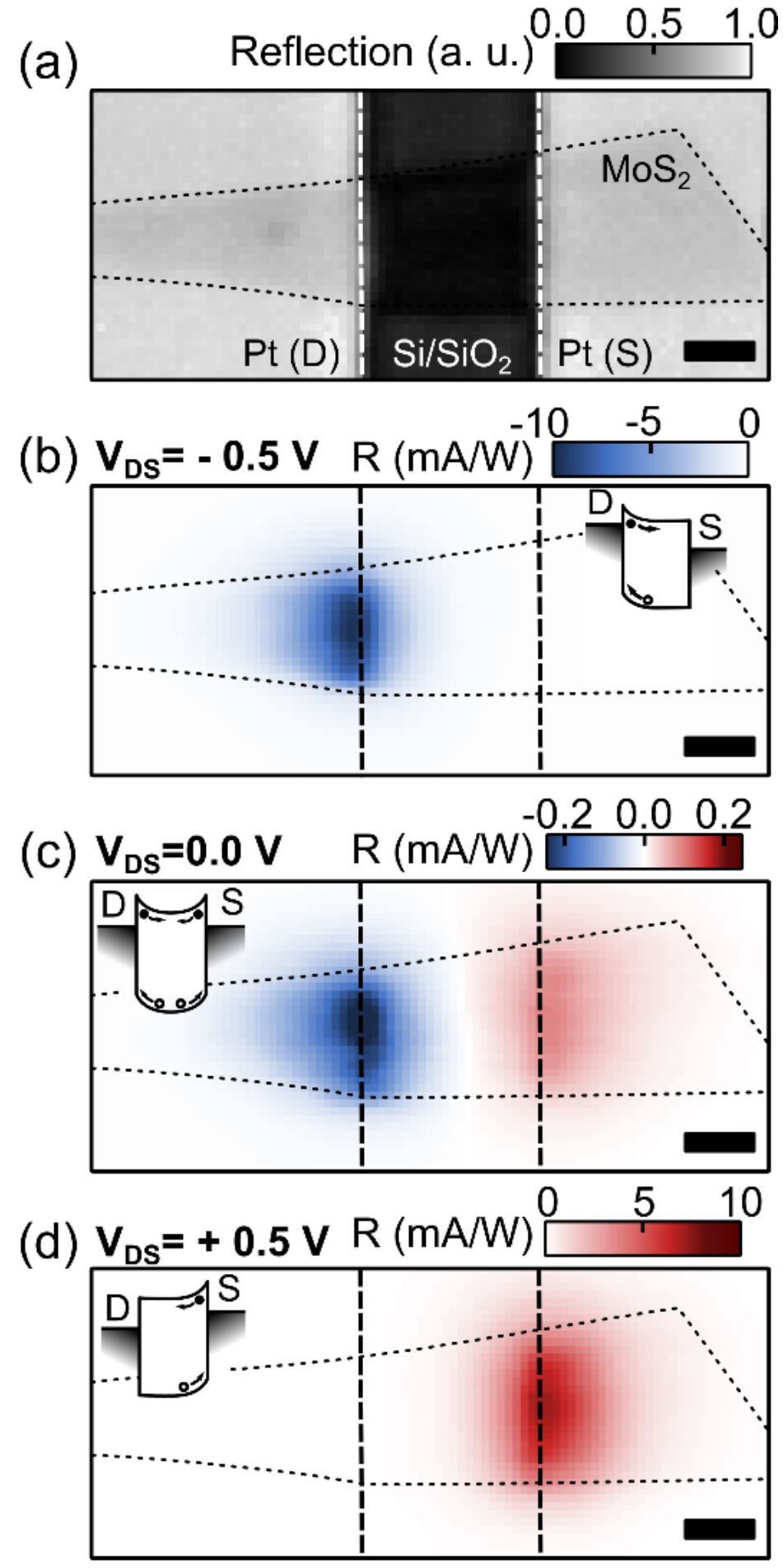

*Figure 3.* Spatially resolved reflection and photocurrent map on a four-layer $MoS_2$ flake. (a) Gray scale reflection map of the device area under investigation. Materials and regions are indicated in the figure. (b-d) Photocurrent maps acquired under different source-drain ($V_{DS}$) potentials. The dashed outlines mark material boundaries, referenced from the reflection map in (a). Inset schematics show the band configuration at each. The scale bar corresponds to 2 µm. A red

single-mode fiber-coupled laser and a microscope 50 × objective (650 nm, 90μW) was employed, achieving a spot size of 1 μm.

Additionally, to further evaluate the versatility of our home-built SPCM system, we employed high-power LEDs coupled with a multimode optical fiber (50 μm diameter) in the same device that was analysed in Figure 3, obtaining spot sizes around 1 μm (see Figure S5). Reflection and photocurrent maps were acquired at 0 V using three LED illumination wavelengths: 625 nm, 530 nm and 470 nm (Figure 4). The reflection maps (Figure 4a-c) clearly distinguish among the Pt electrodes, Si/$SiO_2$ substrate and the $MoS_2$ flake. For the three different wavelengths, in the reflection maps, it can be noticed that Si/$SiO_2$ absorbs more than $MoS_2$ and Pt, and Pt reflects more. Notably, the reflection of the $MoS_2$ flake shows the greater differences, as the blue-green wavelengths exhibit lower reflection, consistent with higher absorption by $MoS_2$ in this spectral region[47,48]. This wavelength dependence is also observed in the corresponding photocurrent measurements at 0 V. Additionally, the reflection maps evidence some drifts during the scan, highlighting the importance of simultaneously recording the reflection image during the scanning photocurrent measurements, in order to properly assign the position of the photocurrent within the spatial map.

The corresponding photocurrent maps were normalised to the incident power of each LED to obtain the responsivity ($R=I_{ph}/P$) of the device, enabling a better comparison between the different light sources (Figures 4 (d-f)). As expected, responsivity is higher for the blue LED, in line with the increased absorbance of $MoS_2$ at shorter wavelengths, as also reported in the reflection maps, corroborating the correlation between the reflection data and the absorbance spectra of the four-layer $MoS_2$ flake[48].

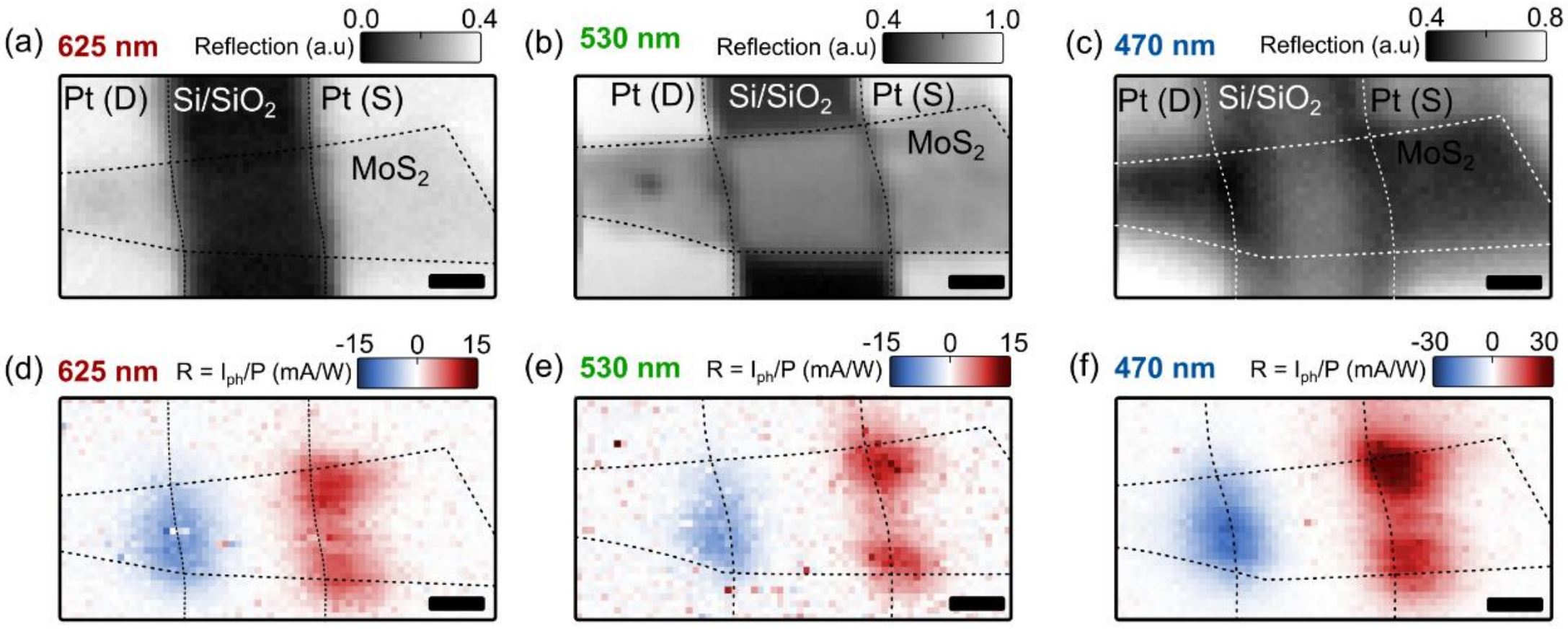

*Figure 4.* Spatially resolved reflection and photocurrent map on a four-layer $MoS_2$ flake. (a-c) Reflection maps of the device using LED illumination at 625 nm, 530 nm and 455 nm, respectively. The different regions and materials (Pt, $Si/SiO_2$ and $MoS_2$) are labelled and defined by a dashed line in each map. (d-f) Corresponding photocurrent maps acquired under zero source-drain bias (0 V) for each illumination. The dashed outlines mark material boundaries, referenced from the reflection maps. Scale bars corresponds to 2 µm

Overall, despite the relatively small flake dimensions and narrow 5 µm channel, our SPCM setup reached sufficient resolution to clearly distinguish spatial variations in photocurrent, and to capture fine features associated with bias-dependent modulation of Schottky barrier, local field effects and wavelength dependency. Achieving high spatial resolution, even when using multimode fiber coupled sources, widens up the range of potential light sources.

Additionally, we fabricated a short-channel device (1 µm channel) by depositing a few-layer graphene (Gr) flake across a pair of commercial Au electrodes (20 µm channel) on a $Si/SiO_2$ substrate, and subsequently cutting the graphene flake using an AFM tip (see Figure 5a). A few layers flake of $WSe_2$ was then transferred to bridge the Gr contacts closing the circuit (Figure 5b).

Again, simultaneous reflection and photocurrent maps of the device were recorded using a high-power red LED (625 nm) coupled to a multimode optical fiber (50 µm core diameter), and scanning at 0.5 µm steps in both the X and Y directions. In the reflection map (Figure 5c), different regions of the scanned area are distinguishable. The Au electrodes exhibit higher reflection values than the $Si/SiO_2$ substrate, even when covered with the Gr flake, which is almost transparent to the light-source employed. The $WSe_2$ flake is clearly visible in the middle of the $Si/SiO_2$ channel, as indicated in the figure.

The photocurrent map recorded at a bias potential of +0.1 V (Figure 5d) shows a localized photocurrent response at the centre of the $WSe_2$ flake, corresponding to the narrow channel between the two Gr electrodes. This result further demonstrates the high spatial resolution of our home-built SPCM setup, capable of resolving photocurrent generation in microchannels.

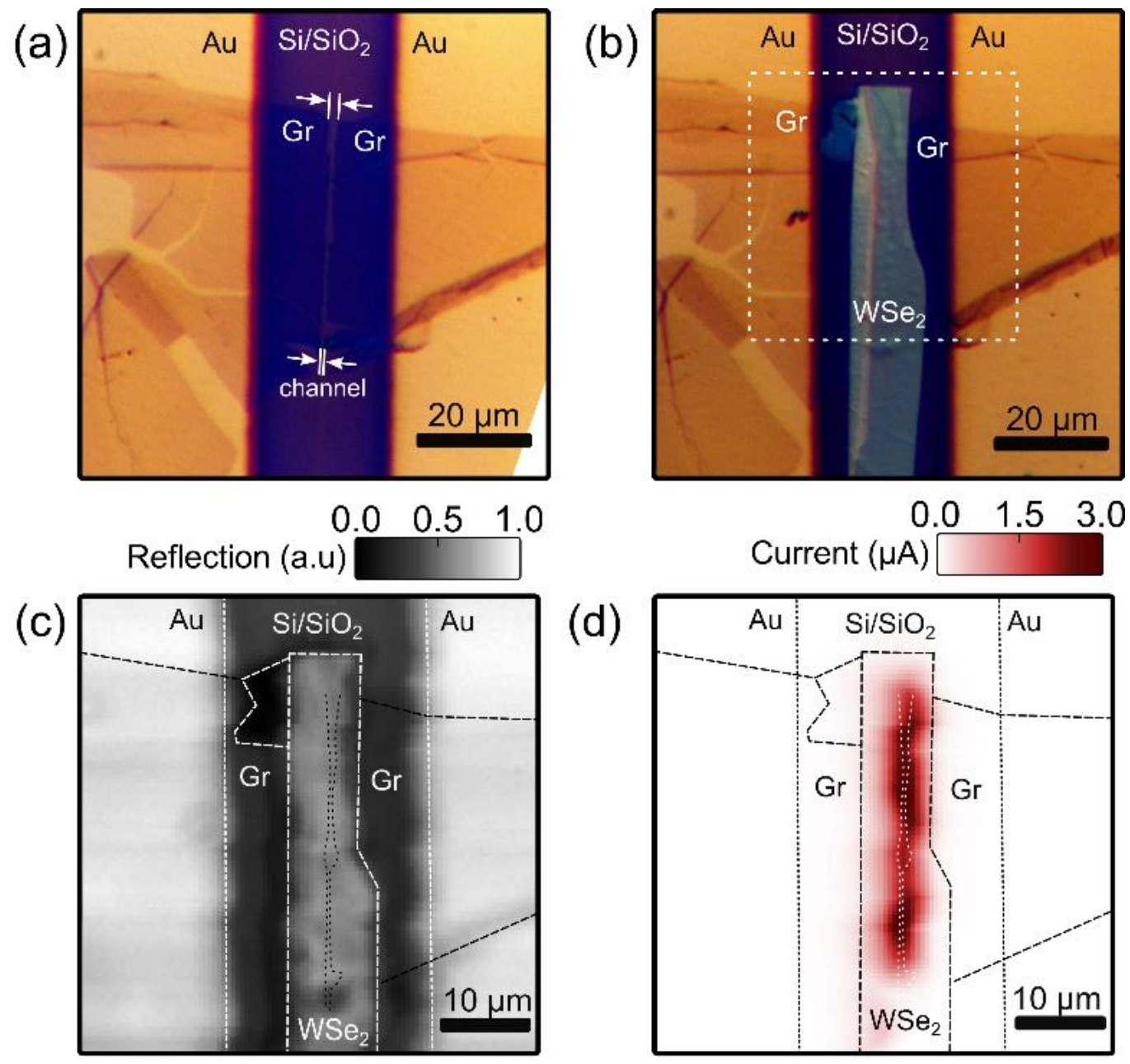

Figure 5. (a) Optical microscopy image of the graphene (Gr) flake transferred onto a pair of Au electrodes on $Si/SiO_2$ substrate after channel definition by AFM tip cutting. The channel between the Gr contacts is indicated in the image with a white dashed line. (b) Optical image of the device after transferring a few-layer $WSe_2$ flake to the bridge the gap between the two Gr electrodes. Scanned region with the SPCM microscope is indicated with a white dashed square. (c) Reflection map showing clear contrast between the Au electrodes, the $Si/SiO_2$ substrate, the Gr regions, and the $WSe_2$ flake as indicated in the map. (d) Photocurrent map recorded at a bias voltage of +0.1 V. Dashed lines limit the different areas as indicated in (c).

## Conclusion

In this work we developed a versatile and cost-effective scanning photocurrent microscopy (SPCM) system capable of resolving sub-micron optoelectronic features in 2D materials and devices. The integration of fiber-coupled light-sources, a motorized XY stage, a digital camera, and an electrical readout module into a standard metallographic microscope, enables tunability of the setup while offering performances comparable to commercial SPCM setups.

The modular design of the setup is compatible with single-mode lasers, multimode LEDs, multiple fiber diameters and standard objective lenses, allowing users to easily adapt the illumination scheme, spectral range, and spatial resolution to the needs of the experiment. The reproducibility of the presented setup is further supported by a detailed calibration procedure, operation steps and a list of the optic, mechanic and electronic components needed.

Through systematic tests on $MoS_2$, $WSe_2$ and InSe devices, we demonstrated that the system can accurately map the reflection contrast and photocurrent response at different bias potentials, as well as wavelength-dependent responsivities even in confined geometries and heterogeneous flakes configurations.

Overall, this home-built SPCM platform offers a robust, customizable and highly reproducible tool for spatially resolved optoelectronic characterization of 2D devices at the microscale, enabling broader adoption of SPCM technique across different laboratories.

## Experimental Section

### Sample Fabrication

InSe (see [49]), $MoS_2$ (Molly Hill mine), $WSe_2$ (HQ Graphene) and Gr nanoflakes (Flake Graphite from NGS) were mechanically exfoliated from bulk crystal using scotch tape (MagicTape) and Nitto tape (SPV 224). The exfoliated nanoflakes were then transferred onto a polydimethylsiloxane (PDMS) Gel-Film substrate (WF ×4 6.0 mil), and their uniformity on the PDMS Gel-Film was verified using an optical microscope in transmission mode. For device fabrication, the $MoS_2$ and Gr/ $WSe_2$ flakes were transferred from the PDMS to a pre-patterned Pt electrode (Ossila S403A2) and Cr/Au electrodes (deposited by thermal evaporator with a custom-made shadow mask in brass, 5nm/45nm), respectively. The thickness of the Pt film is approximately 40 nm, and of the Au film is approximately 50 nm. The fabrication of the electrodes in a compact disk (CD) and the deposition of a $MoS_2$ flake can be found in the Supporting Information-section4.

### Reflection images, AFM and optical microscopy characterizations

Optical images of the flakes, patterned electrodes and devices were captured using an optical microscope (Motic BA310 MET-T). The thickness of the InSe flake and Pt electrodes were measured using atomic force microscopy (AFM, Bruker, Dimension Icon). Laser/ LED reflection intensity images were obtained using the home-built scanning photocurrent microscopy system, equipped with a digital camera, providing spatially resolved reflection images. Further details of the reflection images acquisition and light sources can be found in the Supporting Information (Section 2).

### Optoelectronic Characterizations

Optoelectronic characterizations of the Pt-$MoS_2$-Pt, Gr-$WSe_2$-Gr and Au-$MoS_2$-Au (on CD) devices was performed in atmospheric conditions at room temperature using the home-built scanning photocurrent microscopy system. Further details of the operation of the SPCM, spot sizes, power and calibration can be found in the Supporting Information-section 2 [50].

## Author contributions

**Nuria Jiménez-Arévalo** and **Dan Zheng** contributed equally to this work. **Nuria Jiménez-Arévalo** performed the majority of the experimental work in Madrid, including microscope assembly, sample fabrication and measurements, calibration, data analysis, software development support, and preparation of the first draft of the manuscript. **Dan Zheng** performed equivalent experimental implementation and measurements in Xi'an, including equipment construction, data acquisition, and investigation. Parallel experimental platforms were established and operated at both locations. **Yong Xie** contributed to software development, data processing, and experimental support and supervision. **Rafael Luque Merino** fabricated the narrow-channel graphite electrode devices. **Peng Cheng** contributed to equipment construction, data curation, and investigation. **Yi Liu**, and **Haiwen Yu** contributed to formal analysis, validation, and supervision. **Tao Wang** co-led the project and contributed to conceptualization and funding acquisition. **Riccardo Frisenda** contributed to the design and assembly of the experimental setup and led the development of the control and acquisition software. **Qinghua Zhao** contributed to the early implementation of the microscope, fabrication and measurement of the first generation of samples, and later co-led the project, including conceptualization and funding acquisition. **Andres Castellanos-Gomez** co-led the work, conceived and designed the system, supervised the research, supported implementation and analysis, and contributed to funding acquisition and manuscript preparation.

All authors discussed the results and contributed to the final manuscript.

## Conflicts of interest

The authors declare no competing interests.

## Acknowledgments

This work was supported by National Natural Science Foundation of China (52302199, 52072300, 52472014), the Fundamental Research Funds for the Central Universities, the Research Fund of the State Key Laboratory of Solidification Processing (NPU), China (Grant No.2024-TS-04). A.C-G. acknowledges support from Grants PDC2023-145920-I00 and PID2023-151946OB-I00, funded by MICIU/AEI/10.13039/501100011033 and, respectively, by the European Union NextGenerationEU/PRTR (PDC2023-145920-I00) and by ERDF/EU (PID2023-151946OB-I00). A.C-G. also acknowledge funding from the European Research Council (ERC) through the ERC-PoC 2024 StEnSo project (grant agreement 101185235) and the ERC-2024 SyG SKIN2DTRONICS project (grant agreement 101167218). N.J.A. acknowledges support from the Juan de la Cierva Grant JDC2023-052025-I, funded by MICIU/AEI/10.13039/501100011033 and by ESF+. ICMM-CSIC authors acknowledge support from the Severo Ochoa Centres of Excellence program through Grant CEX2024-001445-S, funded by MICIU/AEI/10.13039/501100011033.

# Modular and Cost-effective Scanning Photocurrent Microscopy System for Sub-micron characterization of 2D optoelectronic devices

Nuria Jiménez-Arévalo [1,†,*], Dan Zheng [2,3,4; †], Yong Xie [1], Peng Cheng [2,3,4], Yi Liu [2,3,4], Rafael Luque Merino [1], Haiwen Yu [2,3,4], Tao Wang [2,3,4], Riccardo Frisenda [5], Andres Castellanos-Gomez [1,*], Qinghua Zhao [2,3,4,*],

1. 2D Foundry Research Group, Instituto de Ciencia de Materiales de Madrid (ICMM-CSIC), Madrid, 28049, Spain
2. State Key Laboratory of Solidification Processing, Northwestern Polytechnical University, Xi'an, 710072, P. R. China
3. Key Laboratory of Radiation Detection Materials and Devices, Ministry of Industry and Information Technology, Xi'an, 710072, P. R. China
4. Research &Development Institute of Northwestern Polytechnical University in Shenzhen, Shenzhen, 518063, P. R. China
5. Dipartimento di Fisica, Università di Roma "La Sapienza", 00185, Rome, Italy

## Supporting Information

### Section 1- Further details of the components and the experimental setup

Table S1 gives a detail description, part number and distributor of the different pieces needed to assemble the scanning photocurrent mapping setup. Figures S1 (a) and (b) detail the commercial parts (with model information) used for fiber input and camera fixture assembly.

**Table S1.** Components of the scanning photocurrent microscopy system indicating the part number, the distributor, and the quantity.

| Description | Distributor | Part number | Quantities |
|---|---|---|---|
| Nexus Optical Breadboard, 24" x 36" x 2.4", 1/4"-20 Mounting Holes | Thorlabs | B2436F | 1 |
| Microscope | Motic | BA310MET-H | 1 |
| Motorized XY Scanning Stage | Standa | 8MTF-102LS05 | 1 |
| Stepper & DC motor controller | Standa | 8SMC5-USB-B9-2 | 1 |
| Joystick | Standa | 8JXY-03 | 1 |
| Power Supply for the motorized XY stage | Standa | PS36-4.4-4 | 1 |
| Clamping Fork | Thorlabs | CF125C/M-P5 | 1 |
| Ø25.0 mm Pillar Post | Thorlabs | RS300/M | 1 |
| XY Stage with Ø1" Hole | Thorlabs | XYR1/M | 1 |
| Coupler External Threads 0.5” long | Thorlabs | SM1T2 | 2 |
| Coupler External Threads 1” long | Thorlabs | SM1T10 | 1 |
| 30 mm Cage Plate with Removable Filter Holder | Thorlabs | CFH2R/M | 1 |
| Adapter with External C-Mount Threads and External SM1 Thread | Thorlabs | SM1A39 | 1 |
| XY Translator with micrometer drives | Thorlabs | ST1XY-S | 1 |
| SM1 Zoom Housing | Thorlabs | SM1NR1 | 1 |
| FC/PC Fiber Adapter Cap with Internal SM1 | Thorlabs | S120-FC | 1 |
| SMA Fiber Adapter Plate with External SM1 | Thorlabs | SM1SMA | 1 |
| Adjustable Lens Tube 0.31” | Thorlabs | SM1V05 | 1 |
| Adjustable Lens Tube 0.81” | Thorlabs | SM1V10 | 1 |
| 10:90 (R:T) Non- polarizing Beamsplitter cube | Thorlabs | BS043 | 1 |
| Compact Clamping 4-Port Prism/Mirror 30 mm Cage Cube | Thorlabs | CM1-4ER/M | 1 |
| Source measurement Unit (SMU) | Keithley | Keithley 2450 | 1 |
| Laser Source 650 nm, 27 mW | Optogear | Optogear 650nm | 1 |
| High Power LED 625nm, 13.2 mW | Thorlabs | M625F2 | 1 |
| High Power LED 530nm, 6.8 mW | Thorlabs | M530F2 | 1 |
| High Power LED 455nm, 9.5 mW | Thorlabs | M455F1 | 1 |
| C-mount Surveillance digital camera. USB Interface | | | 1 |

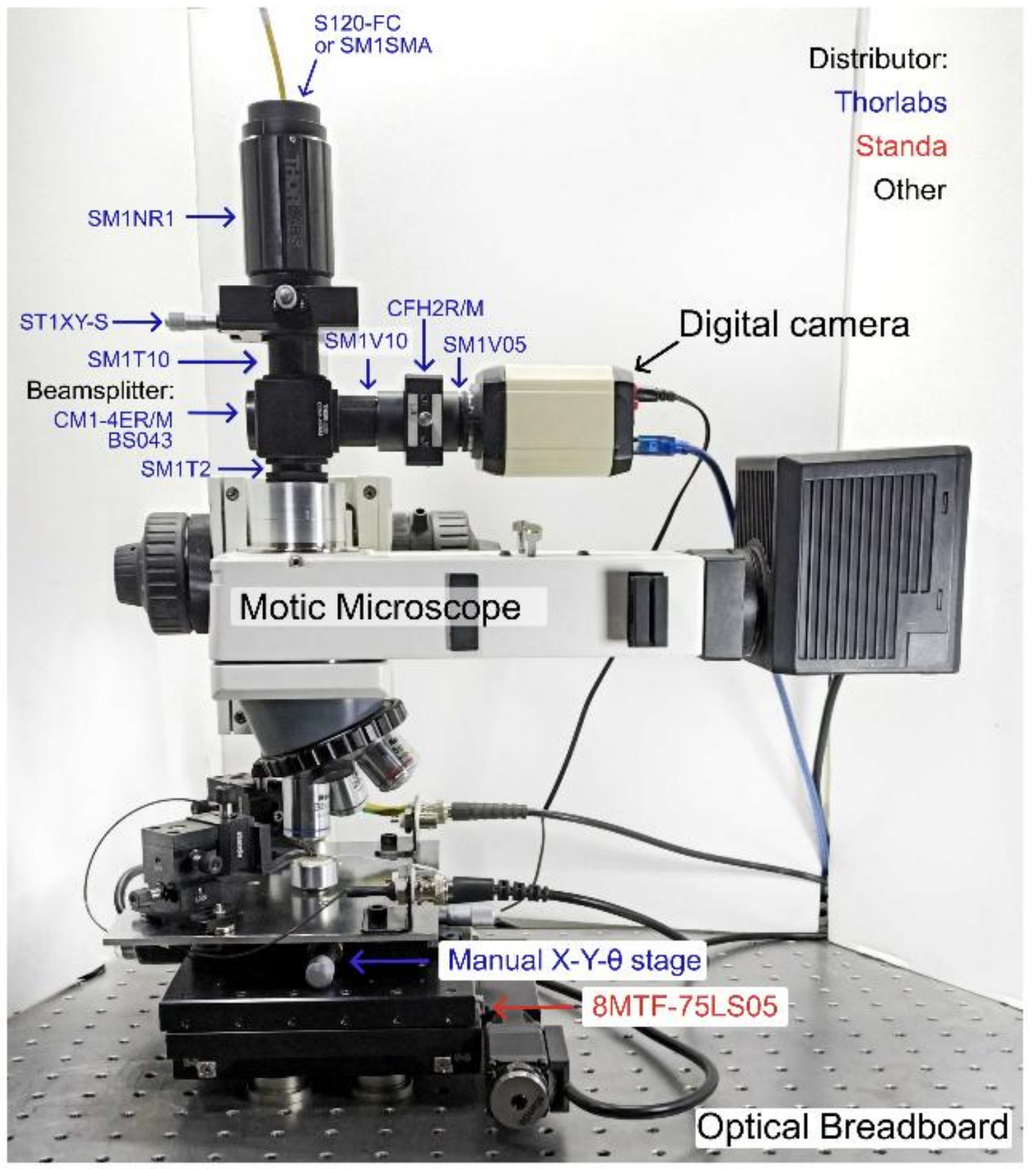

*Figure S1.* Photograph of the Scanning Photocurrent Microscope setup. Main components after the modification of the conventional components are indicated in the image.

## Electrical probes

Figure S2 shows further photographs of the electrical probes employed in this setup. The components needed are detailed in table S2 and the 3D printed electrical cantilever contact is given in [1].

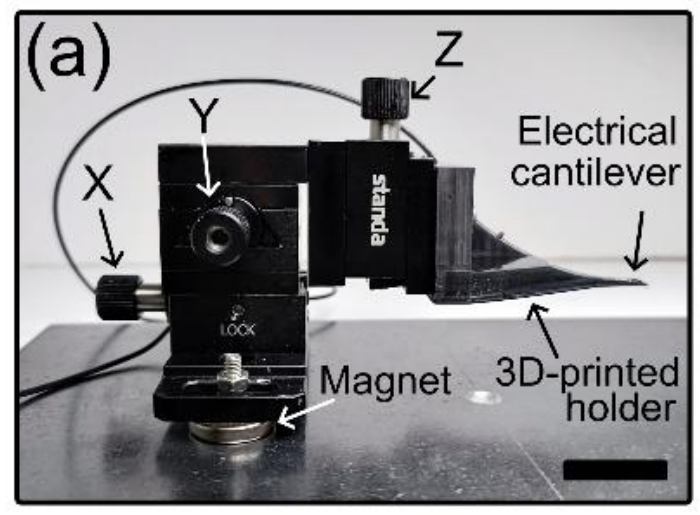

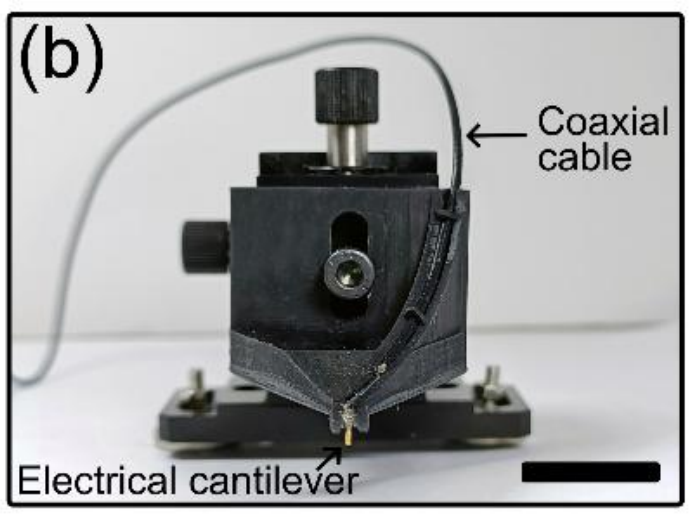

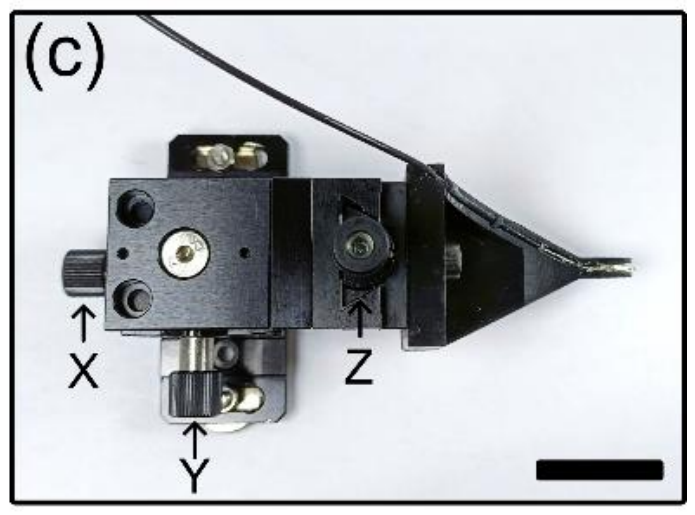

*Figure S2.* Photographs of the electrical probes assembly employed in this SPCM setup, with key components labeled. (a) Side view, (b) front view, (c) top view. Scale bar corresponds to 2 cm. The design files for the 3D-printed holder for the electrical cantilever are available in [1].

**Table S2.** Components of the sample stage and electrical probes indicating the part number, the distributor, and the quantity.

| Description | Distributor | Part number | Quantities |
|---|---|---|---|
| Stainless steel Sample stage | Customized | - | 1 |
| XYZ Dovetail Micromanipulators | Standa | 7T25-10XYZ | 2 |
| Electrical cantilever holder | Customized | -- | 2 |
| Magnets | -- | -- | 4 |
| Electrical cantilever | Molex | 503500-0991 | 2 |
| Coaxial cable | -- | -- | |
| Coaxial connectors RS PRO | RS | Coaxial connector RS PRO | 2 |
| Circular ceramic heating plate | Aliexpress | MDCH Round Heating Plate 48MM | 1 |

## Section 2- Further details of the SPCM setup

### Construction of the reflection map using an imaging digital camera

In the present SPCM system, the reflection map is directly obtained from the microscope camera during the photocurrent scan, without the need for a dedicated photodetector.

The digital camera images the illuminated region of the sample through the 10% reflection path of the beamsplitter.

During alignment, white light illumination is used to visualize the device geometry and to position the focused optical spot. At this stage, the laser or LED is switched on at minimum power just sufficient to identify the position of the spot, and the area to scan is defined (see Figures S3 a,c). Once the spot position is optimized, white light is switched OFF, the power of the laser or LED is increased, and a neutral density (ND) filter is placed in front of the camera to prevent saturation during high-power illumination.

Under these conditions, the camera records the spatial intensity distribution of the illumination spot reflected from the sample surface. Using a custom-developed graphic user interface created for the SPCM system, the pixels of the recorded image corresponding to the laser the laser/LED spot are defined, and the reflected intensity of the sample is scanned.

During the SPCM scan, an image is acquired at every scan position. For each frame, the reflection signal is quantified by selecting a color channel that matched the illumination wavelength (red, green or blue) over the defined region that corresponds to the illumination spot. This scalar intensity value of the reflection is then assigned to the corresponding scan coordinate, producing a two-dimensional reflection map that is registered simultaneously with the photocurrent data.

As shown in Figures S3b and S3d, the reflected intensity exhibits a clear contrast between Au and $SiO_2$ regions, arising from their different optical reflectivity at the illumination wavelength. This contrast is observed for both coherent laser illumination (650 nm) and incoherent LED illumination (625 nm), and it is used to create the reflection map in which the different regions can be distinguished.

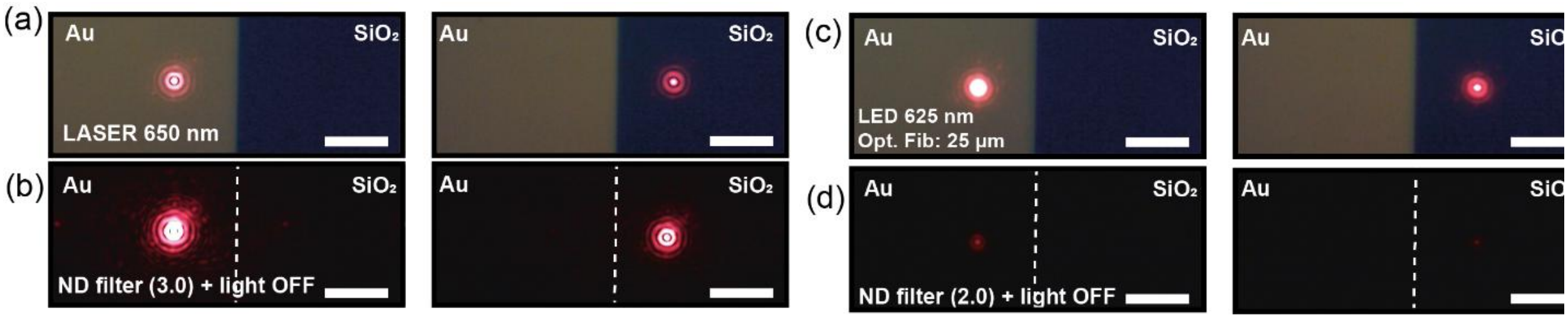

*Figure S3.* (a) Laser illumination (650 nm) with white light ON and low laser intensity, showing the spot centered on Au and $SiO_2$ surfaces. (b) Same laser source at maximum power with white light OFF and an ND filter (3.9). (c) LED illumination (625 nm) using a 25 μm optical fiber, with white light ON and LED at maximum intensity with the spot centered on Au and $SiO_2$ surfaces. (d) same LED configuration with white light OFF and an ND filter (2.0) showing the LED spot on the Au and $SiO_2$. Dashed lines in (b) and (d) mark the interfaces between Au and $SiO_2$. Scale bars: 10 μm.

## XY stage calibration

To calibrate the step displacement of the motorized XY platform, a high-resolution digital indicator (AUTOUTLET) was employed and carefully aligned with the XY stage such that its sensitive axis was parallel to the motion direction. The stage was displaced by a known number of motor steps, and the resulting linear displacement was measured using the digital probe. By plotting the measured displacement as a function of the number of commanded steps and fitting a linear function (Figure S4), a calibration factor of 1.25 μm per full step was obtained for both the X and Y directions.

The motorized stage controller allows operation in micro-stepping mode, in which each full step is subdivided electronically. Under the micro-stepping configuration used in this work, this corresponds to a minimum incremental displacement of 0.125 μm per micro-step, enabling sub-micrometre positioning accuracy.

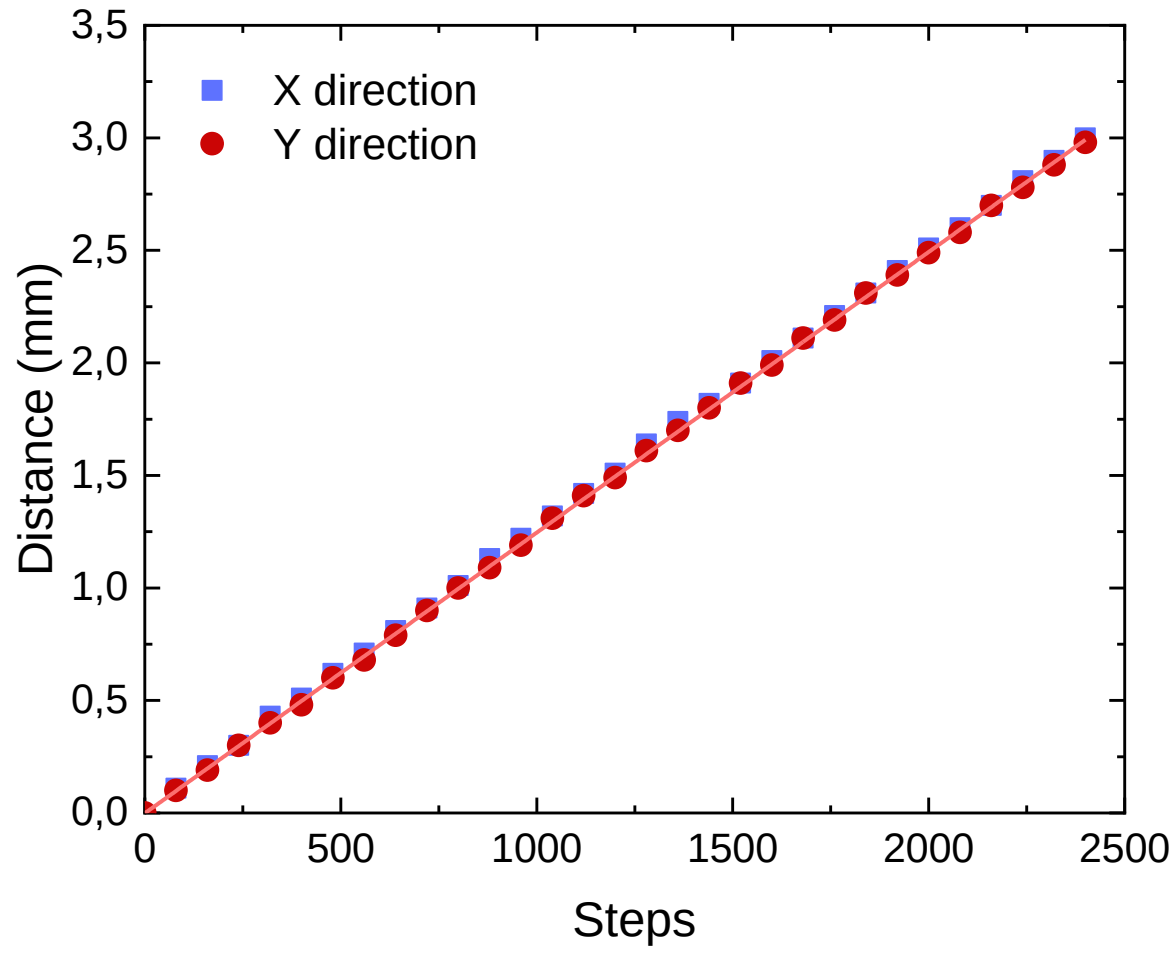

*Figure S4*. X-Y stage calibration curve. Linear fit to the Y direction, showing a slope of 1.25 ± 0.01 μm/step.

**Estimation of the Laser Spot Size**

The spot size was determined by scanning in reflectance through an abrupt edge, in this case created by the cleaved edge of a gold-coated $SiO_2$/Si wafer. This configuration provides a sharp transition from a highly reflective surface to an air gap, ensuring maximum contrast in the reflected signal, essential for accurate spot size estimation (see Figure S5a). Alternatively, this calibration can also be done using the interface between a gold electrode and a bare $SiO_2$/Si substrate, although contrast and sharpness are slightly lower (see Figure S3).

During the scan, the intensity of reflection at each point along a line crossing the interface was recorded. The resulting signal resembles of a smoothed step function, which corresponds to the convolution of a step function with the Gaussian intensity distribution of the laser point. To model the measured profile, we used the following fitting function.

$$f(x)=y_0+\frac{h}{2}\cdot erf\left(\frac{\xi}{w}\right)+\alpha\cdot\xi+\frac{\beta}{2}\cdot(|\xi|+w\cdot exp\left(-\frac{|\xi|}{w}\right); \qquad \xi=x-x_0 \qquad (eq.1)$$

Here, $y_0$ accounts for the background level, $h$ is the amplitude of the reflectance change, $x_0$ is the center of the transition, and $w$ characterizes the width of the Gaussian beam profile. The additional linear $(\alpha\cdot\xi)$ and the exponential tail ($\beta$-dependent) terms account for background slope and edge diffusion effects, respectively.

Although the function includes additional correction terms, the parameter $w$ appears in the argument of the error function in the form $erf\left(\frac{\xi}{w}\right)$, which implies a relation to the standard deviation of a Gaussian beam $\sigma=w/\sqrt{2}$. Therefore, the full width at half maximum (FWHM) of the laser spot is given by:

$$FWHM=2\sqrt{2ln2}\cdot\sigma=2\sqrt{ln2}\cdot\sigma\approx 1.665\cdot w \qquad (eq.2)$$

We used equation 2 to extract the laser spot size from the fitted parameter $w$, under different illumination conditions. Initially, a single-mode fiber-coupled diode lasers (650 nm) was tested (Figure S5a). With a 50× microscope objective (NA=0.55), we observed that the system achieved a spot size of 1 μm close to the Rayleigh diffraction limit for that wavelength (721 nm).

We also explored the use of multimode optical fibers to increase versatility (Figure S5b). Although multimode fibers can lead to broader spot profiles due to their larger core diameter and mode mixing, they offer the advantage of supporting a broader range of light sources, including fiber-coupled high-power LEDs and tunable light sources, using the same optical path. We tested fibers with 105 μm and 25 μm core diameters. Notably, the combination of high-power LEDs with a 25 μm fiber core, together with 20× or 50× objectives, still yielded diffraction-limited spot sizes and sub-micron spot sizes (0.66 μm and 0.75 μm for the blue and green LEDs, respectively), while maintaining high enough intensity to perform reliable scanning photocurrent measurements across a wide variety of devices (see Figure S5c).

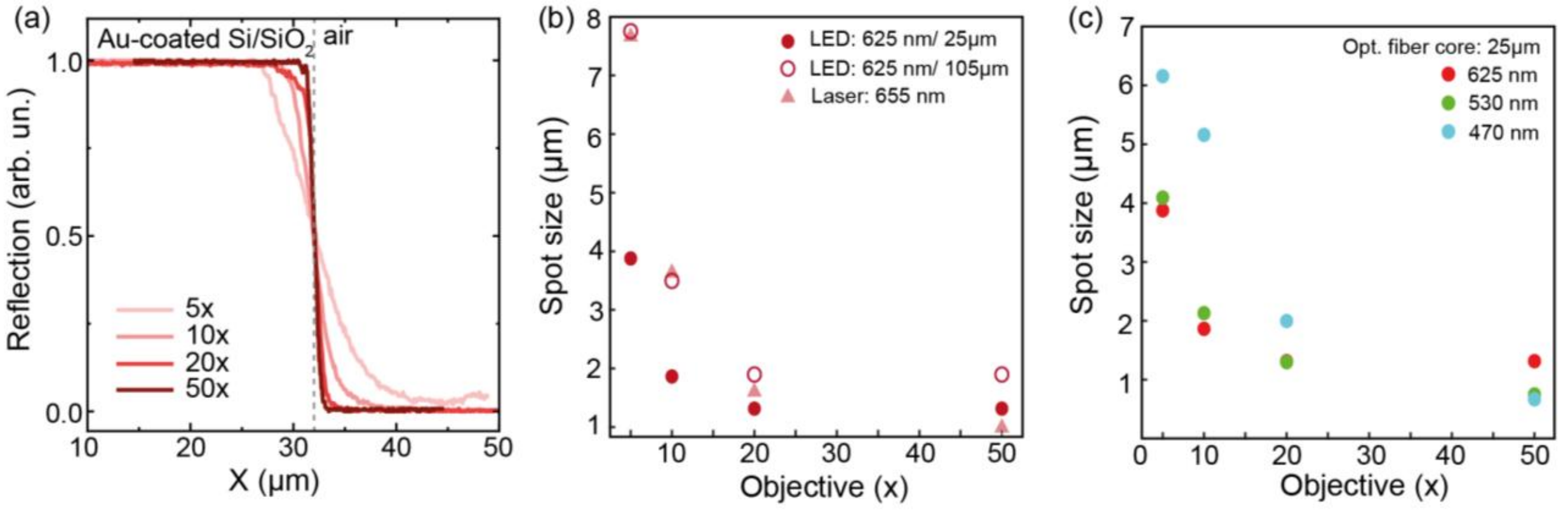

*Figure S5.* (a) Reflection line profile across the edge of a gold-coated Si/$SiO_2$ substrate, measured using a laser of 650 nm and objectives with different magnifications (5×, 10×, 20× and 50×). The dashed line marks the transition from Si/$SiO_2$ to air. (b) Calculated spot sizes for each objective, using different illumination sources (LED and laser) and, for the LED, multimode optical fibers with core diameters of 25 μm and 105 μm, as indicated in the legend. (c) Spot size using the optical fiber with a core of 25 μm, and the illumination using LEDs with different wavelengths (625 nm, 530 nm and 470 nm).

**Power of the different light-sources**

To measure the power of the different light sources (red laser and high-power LEDs) we employed a power meter that we placed at the end of the light beam after finding the focus. The power for the different light sources using the different objectives of the microscope is represented in Figure S6.

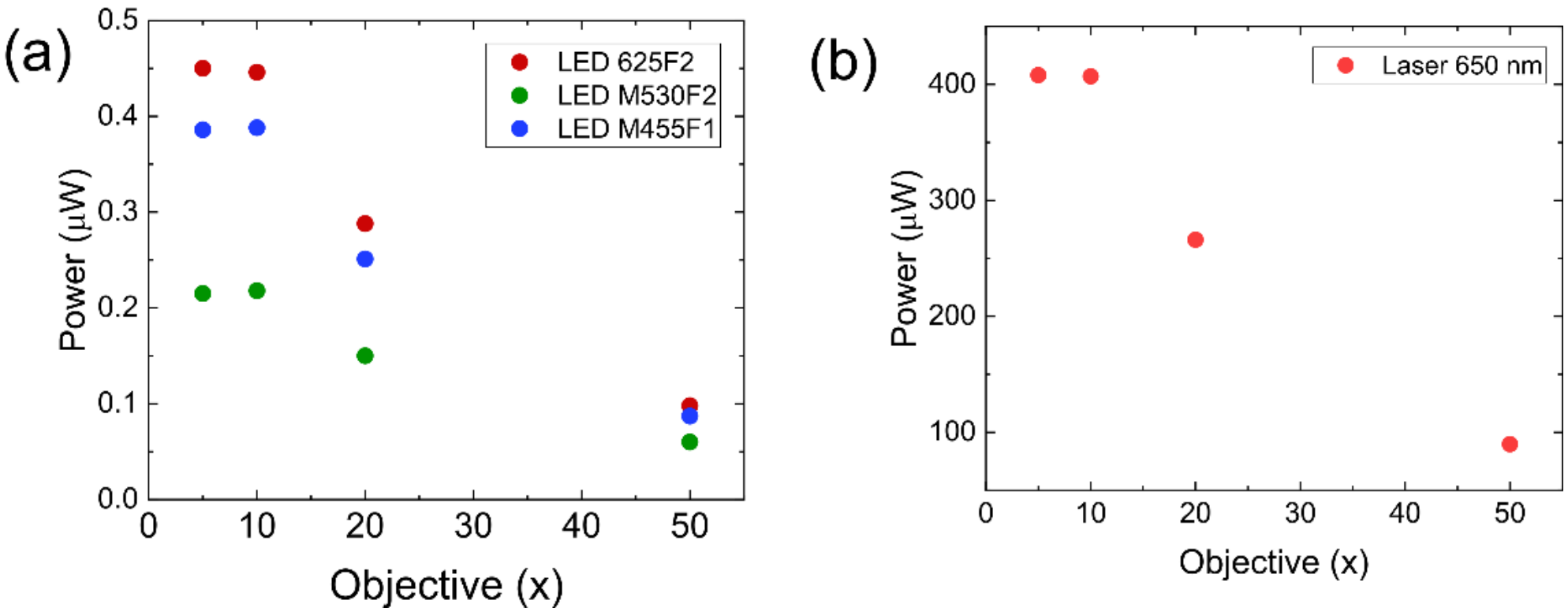

*Figure S6.* Power of the different high-power LEDs using the multimode-optical fiber (50 μm core diameter) (a) and the red laser (650nm) with a single-mode optical fiber (b) as a function of the employed objective.

## Section 3- Experiment Pt-$MoS_2$-Pt

### Morphological and optical characterization

We employed mechanical exfoliation and the deterministic transfer of flakes to fabricate the Pt-$MoS_2$-Pt device on Si/$SiO_2$ that can be seen on Figure S7. The differential reflectance of the flake was acquired before its transfer, consistent with a four-layer thickness.

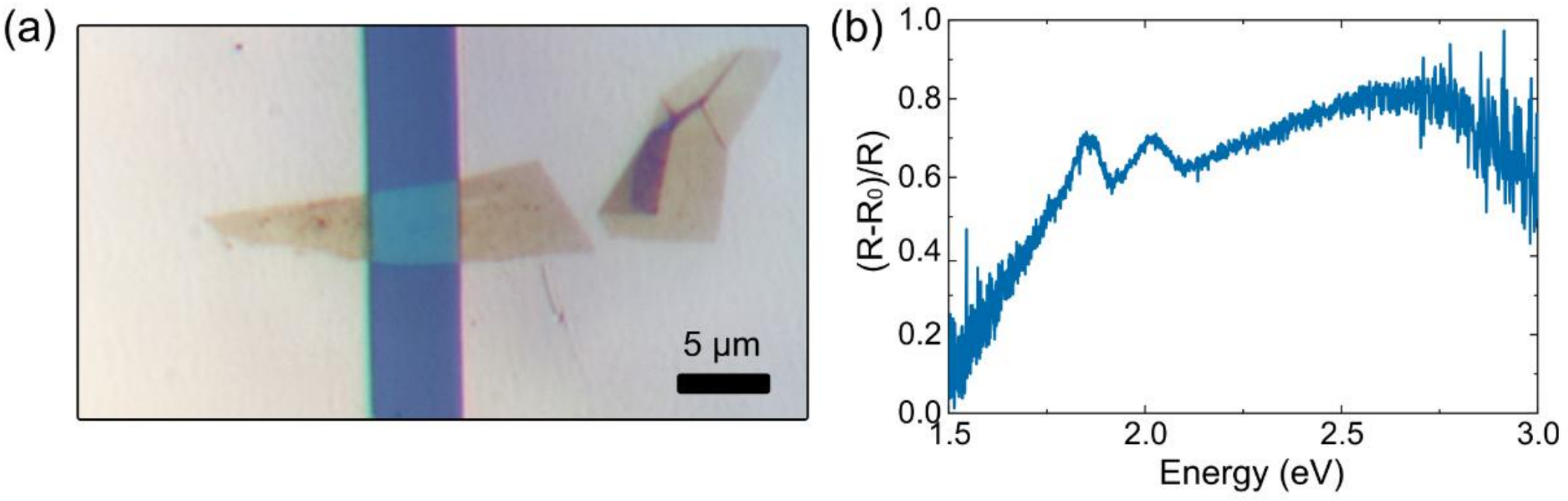

*Figure S7.* (a) Optical microscopy image of the investigated device. The $MoS_2$ flake is positioned within a Si/$SiO_2$ channel, connecting two Pt electrodes. (b) Differential reflectance spectrum of the $MoS_2$ flake, consistent with a four-layer thickness.

## Section 4- Further experiments

### $MoS_2$ on a CD sample

For this experiment, we employed a commercial recordable CD as the substrate. An electrical channel was created using a nail-polish wire as a mask prior to the evaporation of Au. After Au deposition, we removed the nail-polish wire using an ultrasonic bath with ethanol for 2 minutes, resulting in a channel of approximately 15 µm. After that, a $MoS_2$ flake was deposited on the channel, as can be seen in the optical microscopy image in Figure S8a.

Different regions can be distinguished in the image: the Au contacts with the channel in the center, and the flake deposited is in the middle which exhibits multiple regions including a big four-layer (4L) region with a wrinkle in its central part, a multilayer edge in the right side, and distinct cracks in the flake. Overall, we can notice a heterogeneous morphology in this flake comprising several distinct regions. In addition, the CDs exhibit a periodic modulation of 1.6 µm, distance between tracks, which makes this substrate ideal for testing the capability of our SPCM to resolve micrometric architectures.

We simultaneously recorded the reflection and photocurrent signal while scanning the device with the light beam of a high-power red LED (625 nm) coupled to a multimode optical fiber (50 µm core diameter). The scan was performed with 0.5 µm steps in both the X and Y axis. Figure S8b shows the reflection image. We can perfectly distinguish between all the previously mentioned regions, Au, bare CD, $MoS_2$-ML, $MoS_2$-4L, wrinkle and flake-cracks. Additionally, the CD's tracks are also evident across the entire reflection map.

By the analysis of different line profiles in the horizontal direction (Figure S8c), we measured the distance between the modulated stripes of the CD. The two representative line profiles shown in Figure S8c exhibit consistent modulation along the horizontal direction, with an average value of (1.5±0.2) µm, in good agreement with the nominal 1.6 µm modulation of standard CDs. This result demonstrates the good spatial resolution of our SPCM, even when using the high-power LEDs as the light source.

Figures S7d-f show the photocurrent maps at three different applied potentials, -0.1 V, 0.0 V, and +0.1 V, respectively. A clear dependance of the photocurrent sign based on the applied potential can be noticed, consistent with the behavior discussed in the main text of the Pt-$MoS_2$-Pt configuration. Furthermore, due to the inhomogeneity among the different regions, the SPCM maps exhibit different current-dominant regions at different potentials. This result highlights the importance of using SPCM when investigating complex morphologies, as it enables the identification of the contribution from each region under different bias conditions.

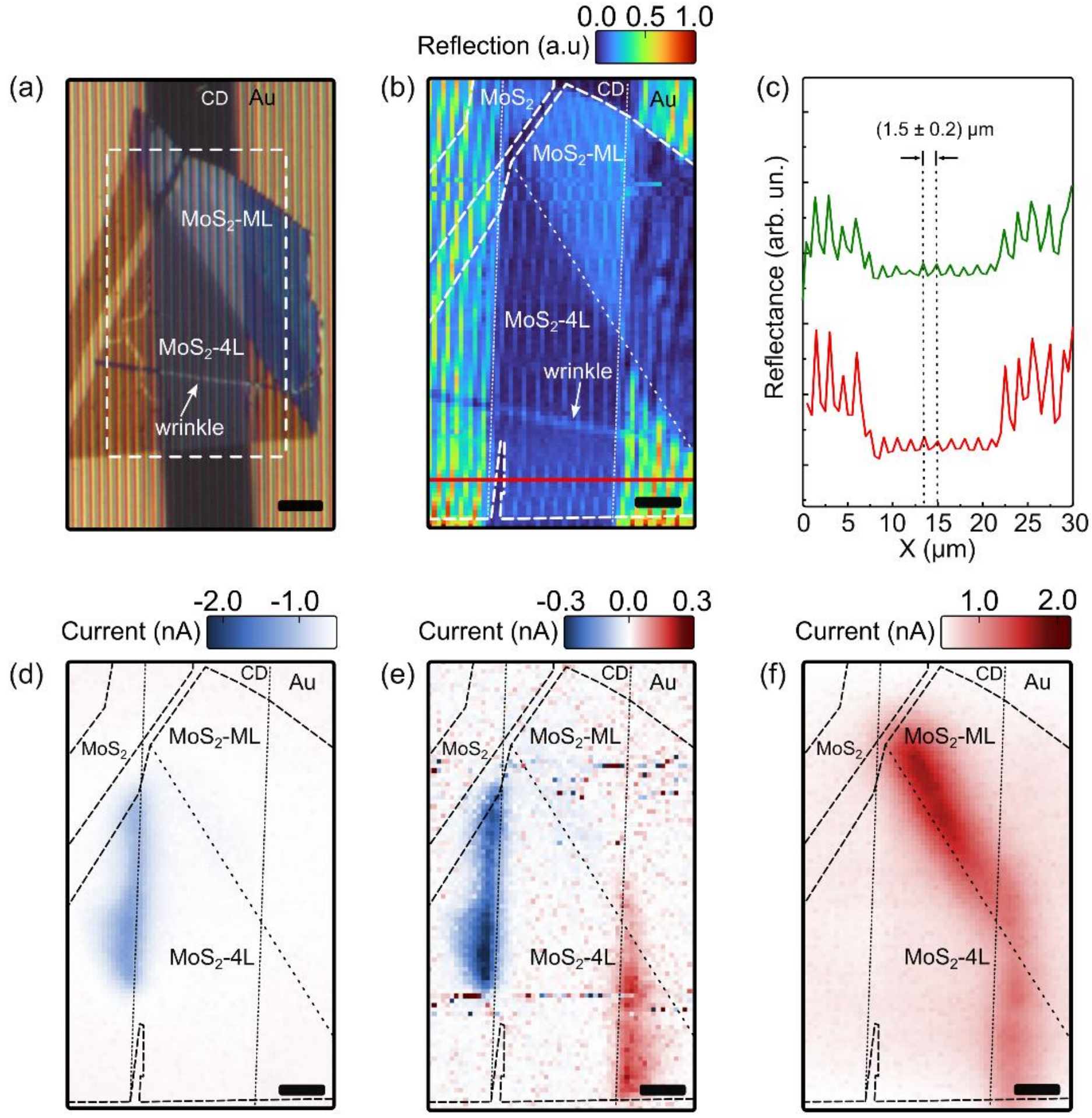

*Figure S8.* (a) Optical microscopy image of the $MoS_2$ flake deposited on the CD substrate. The Au electrodes and the different $MoS_2$ regions (multilayer (ML), four-layer (4L), and a wrinkle) are indicated. The periodic modulation of the CD surface is clearly visible. (b) Reflection map acquired using the SPCM setup, where the modulation of the CD and the contrast between the different $MoS_2$ regions are evident and indicated with dashed lines (c) Representative horizontal line profiles extracted from the reflection map (red one extracted from the horizontal red line in (b)), showing the periodic modulation. (d–f) Photocurrent maps recorded at applied biases of –0.1 V, 0.0 V, and +0.1 V, respectively. The dashed outlines mark material boundaries, referenced from the reflection map in (b). Scale bars: 5 μm.

## Pt-InSe-Gr

To further explore the capabilities of the SPCM system for photoelectric characterization of 2D optoelectronic device, we fabricated a Pt-InSe-Gr (graphite) Schottky device and measured its current-voltage (*I-V*) characteristic using the established system. Figure S9a presents an optical image and measurement schematic of the Pt-InSe-Gr device, fabricated via the dry transfer method. Figure S9b shows the *I-V* curves of the device under both dark and global 530 nm illumination using a green LED coupled to an optical fiber. Global illumination is achieved using a large core multimode optical fiber (400 μm of core diameter) resulting in a spot size on the surface larger than the device dimensions.

The device demonstrates a strong photoresponse, with asymmetric *I-V* curves indicating rectifying behavior[2,3]. As the light power density increases from 6.25 mW/cm$^2$ to 200 mW/cm$^2$, both forward and reverse photocurrents increase within the voltage range of -1 V to 1 V. Under 200 mW/cm$^2$ illumination, the forward current reaches a maximum of 20 nA at 1 V.

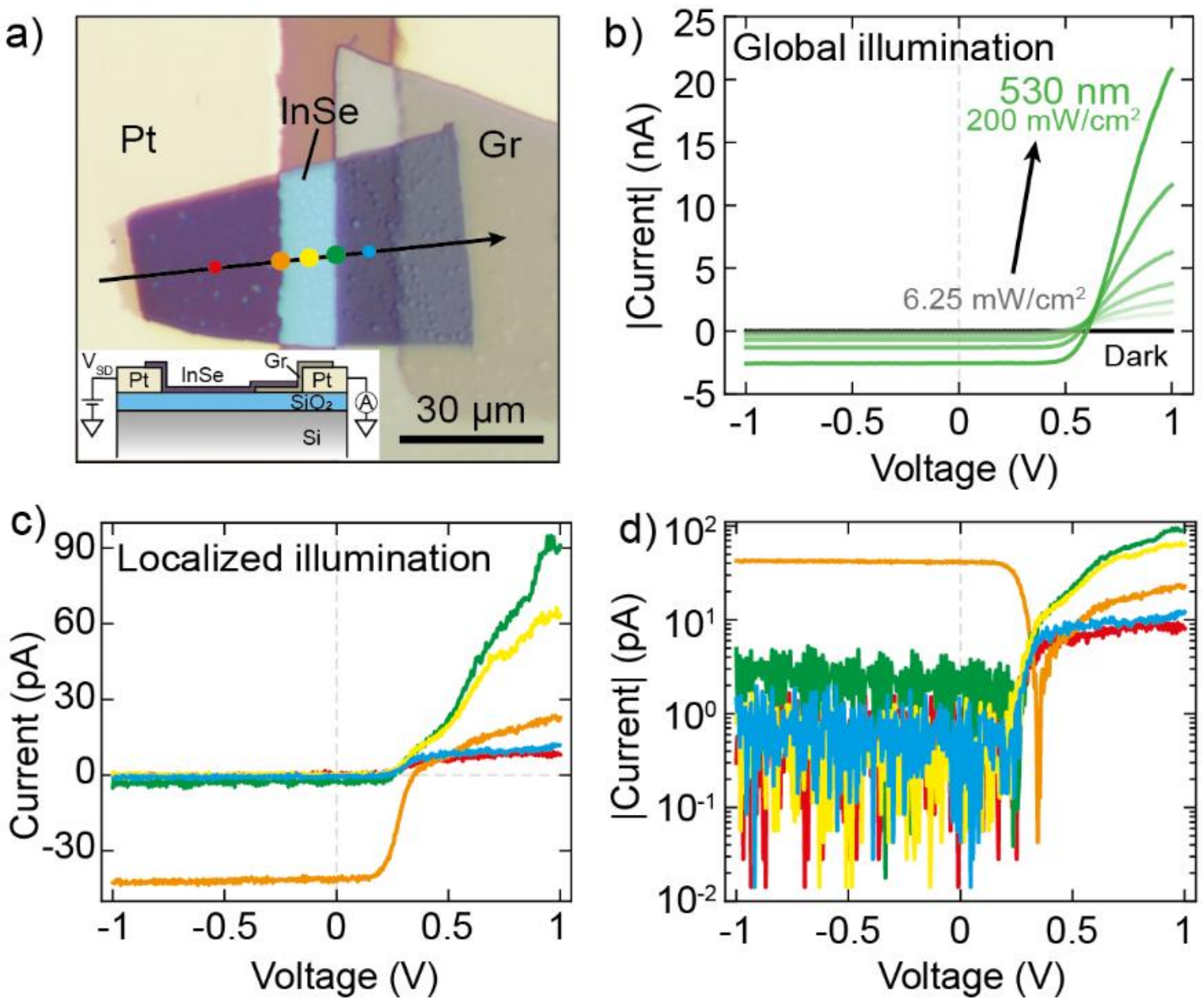

*Figure S9.* Characterizations of a Pt-InSe-Gr device under global and localized illumination. (a) Optical image of the Pt-InSe-Gr device and measurement schematic (inset). (b) *I-V* characteristics recorded under dark conditions and global 530 nm illumination at varying power densities. (c) *I-V* curves in linear scale, obtained by varying the position of the illumination spot on the device. The colors of the curves correspond to the spot positions and colors indicated in panel (a). (d) Semi-logarithmic plot of the *I-V* curves from panel (c), highlighting the open-circuit voltage.

To investigate the photocurrent generation at different positions on the device, we performed *I-V* measurements while illuminating with a single-mode fiber-coupled laser (λ = 660 nm) focused to a small spot (~2 μm) on the device. Figure S9c shows the *I-V* curves recorded at various localized illumination positions, as indicated by the corresponding color spots in Figure S9a. The *I-V* data in Figure S9d are plotted on a semi-logarithmic scale for better visualization. At different bias voltages, very small reverse currents are observed when the laser is focused on the overlap regions, including Pt/InSe (red), InSe/substrate (yellow), InSe/Gr (blue), and the Gr/InSe contact interface (green). However, when the laser is focused on the Pt/InSe interface (orange), the reverse current increases significantly, reaching tens of times the current observed at other positions. At zero external bias ($V$ = 0 V), the *I-V* curve for the Pt/InSe contact interface shows a short-circuit current ($I_{sc}$) of approximately 42 pA and an open-circuit voltage ($V_{oc}$) of around 0.4 V, consistent with typical photovoltaic behavior. The photogenerated electron-hole pairs are separated by an internal electric field even without external bias, which could arise from the Schottky barrier at the Pt/InSe interface. In contrast, the near-zero current at 0 V at the Gr/InSe interface suggests an ohmic or low-barrier contact[2–5]. The *I-V* curves obtained from the SPCM system provide valuable insights into the micro-region characteristics of the asymmetric electrodes, as well as the interfacial built-in electric field of the Schottky contact.